\documentclass{aastex}
\usepackage{amssymb,amsmath}
\usepackage{lineno}
\def\app#1#2{%
        \mathrel{%
                \setbox0=\hbox{$#1\approx$}%
                \setbox2=\hbox{%
                        \rlap{\hbox{$#1\propto$}}%
                        \lower1.1\ht0\box0%
                        }%
                        \raise0.25\ht2\box2%
                        }%
                        }

\usepackage[hyphens]{url}
\usepackage{revsymb} 
\begin{document}
\title{Orbital and Precession Periods in Repeating FRB 20121102A}
\shorttitle{Periods in Repeating FRB 20121102A}
\shortauthors{Katz}
\author{J. I. Katz\altaffilmark{}} 
\affil{Department of Physics and McDonnell Center for the Space Sciences,
Washington University, St. Louis, Mo. 63130}
\email{katz@wuphys.wustl.edu}
\begin{abstract}
\cite{LGLW} reported a 4.605 day period in the repeating FRB 20121102A in
addition to the previously reported 157 day modulation of its activity.
This note suggests that the shorter period is the orbital period of a
mass-transferring star orbiting a black hole, possibly of intermediate mass,
and that the 157 day period is the precession period of an accretion disc
around the black hole.  The mass-losing star must be evolved.
\end{abstract}
\newpage
\section{Introduction}
\citet{LGLW} recently reported the discovery of a 4.605 day period in the
bursts of the repeating FRB 20121102A, previously known \citep{R20,C21} to
have its activity modulated with a period of 157 days.  It has been proposed
\citep{K22b} that the 16.3 day periodic activity modulation of the repeating
FRB 20180916B may be attributed to precession of an accretion disc around a
black hole, with fast radio bursts (FRB) emitted from the accretion funnel
near the direction of the angular momentum axis.  Such a model implies the
presence of a binary companion as the source of the accreted matter and of
the torque that drives the precession, hence the existence of an orbital
period.  This note identifies the newly discovered 4.605 day period of FRB
20121102A as the orbital period of its source.
\section{Comparison to Other Precessing Binaries}
Her X-1 was the first binary discovered to have a superorbital period.  This
was soon attributed \citep{K73} to Newtonian precession of the accretion
disc's axis under the influence of its companion star, in analogy to the
precession of the Moon's orbit under the influence of the Sun's gravity.
A precessing disc in SS 433 \citep{K80,M84} accelerates a subrelativistic
jet along its axis that emits hydrogen lines whose Doppler shifts permit
quantitative measurement of the disc's precession \citep{KAMG}.

\citet{L98} summarized the parameters of X-ray binaries with
well-established superorbital periods attributed to disc precession.  The
ratios of superorbital to orbital periods range from 13 to 22.  The two
periods reported for FRB 20121102A have a ratio of 34, somewhat outside the
range of the other systems, but qualitatively similar, supporting the
identification of its periods as orbital and precession.

Theoretical calculations \citep{K73,K24} of the period ratio if the disc
is treated as a ring at the Roche Circularization Radius (RCR) indicate much
larger values, especially if the binary mass ratio is large.  However, such
calculations ignore the fact that viscosity broadens the ring to a disc
after it forms.  The period ratio depends on the unknown distribution of
mass in the disc \citep{L98}, and is much smaller than for a ring at the RCR
because the outward expansion of the disc to its last stable orbit
\citep{BDKP} increases the torque on it.  The period ratio of 34 observed in
FRB 20121102A is likely consistent with attribution of its 157 day period to
disc precession and of its 4.605 day period to its orbit.
\section{Periodicity {\it vs.\/} Periodic Modulation of Activity}
The bursts of FRB 20121102A are not periodic, despite the presence of
periodic signals in the Phase-folding probability Binomial Analysis (PBA).
Only the envelope of burst activity is periodic, as also observed for the
repeating FRB 20180916B \citep{CHIME,P21,Mck23} with a period of 16.3 d
(there is yet no evidence for another period in this system).  This explains
why the PBA can reveal the periodicities of FRB 20121102A while
periodograms, $\chi^2$, H tests and quadratic mutual information analysis,
that are sensitive to periodicities but not to more subtle temporal
structure, do not \citep{LGLW}.
\section{Discussion}
No repeating FRB has shown periodic bursts and upper limits are constraining
\citep{K22a,N22,D23}.  As a result, it has been proposed that their sources
are not magnetic neutron stars, whose emission would be expected to be
periodic at their rotational periods, but rather the throats of accretion
discs around black holes for which no periodicity is expected \citep{K17}.
The source of mass for accretion might be interstellar gas, in which case
again no periodicity would be expected, but it might alternatively be a
binary companion, in which case both orbital and precessional periods exist
and might be observable.  The viewing angle of the throat of a precessing
disc varies at the precession period, naturally modulating the burst rate
\citep{K22b}.  It is harder to see how the orbital period might be
manifested observationally, but it is known that in SS 433 there are
intercombination frequencies in the orientation of the jet, and hence of the
funnel from which it emerges \citep{KAMG}.

The detection of jitter (about the modulation period) in the times of bursts
from FRB 20180916B with fractional amplitude similar to that of jitter in
SS 433 \citep{K22b} strengthened the case for accretion discs as the sources
of repeating FRB.  The recent discovery \citep{LGLW} of a second period in
FRB 20121102 demonstrates the existence of the second period predicted by
precessing disc models, with a period ratio comparable to those of known
precessing discs in binary X-ray sources \citep{L98}.

In the hypothetical semi-detached mass-transferring binary system with
orbital period 4.605 days the density $\rho$ of the mass-losing star may be
estimated from Kepler's law and approximations \citep{E83} to the size of
Roche lobes.  If the ratio of the mass of the accreting black hole to that
of the mass-losing star $q \gg 1$, the result is
\begin{equation}
	\rho \sim 2 \times 10^{-3}\ \text{g/cm}^3,
\end{equation}
nearly independent of $q$ and independent of the masses if $q$ is specified.
This is about two orders of magnitude less than the density of any main
sequence star, implying an evolved star.  The origin of such a system is a
matter for speculation, but it might be either the remains of a massive
binary one of whose members has collapsed to a black hole or an intermediate
mass black hole that captured, by either tidal dissipation or three-body
interaction, a star.

This model suggests that PBA analysis of the times of bursts from FRB
20180916B may reveal another shorter period in addition to its known 16.3 d
period.  By analogy, such an orbital period might be ${\cal O}
(0.5\,\text{d})$.  The same suggestion applies to any other FRB whose
activity is periodically modulated.
\section*{Data Availability}
This theoretical study generated no new data.

\end{document}